\documentstyle[amstex,12pt,twoside]{article}            
\newlength{\dinwidth}
\newlength{\dinmargin}
\newsymbol\rest 1316         
\newcommand{\RR}{{\Bbb R}}                      

\newcommand{\dx}{d^{\, 3} \mbox{\boldmath $x$}}       

\newcommand{\ssx}{\mbox{\footnotesize{\boldmath $x$}}}
\renewcommand{\sp}{\mbox{{\boldmath $p$}}}

\newcommand{\Oo}{{\cal O}}                      

\newcommand{\AO}{{\cal A}({\cal O})}            
\newcommand{\BO}{{\cal B}({\cal O})}            
\newcommand{\BOp}{{\cal B} ({\cal O}^{\, \prime})}         

\newcommand{\Al}{{\cal A}}                      

\newcommand{\Alu}{\underline{{{\cal A}_{ }}}}         
\newcommand{\AOu}{\underline{{{\cal A}_{ }}} ({\cal O})}

\newcommand{\Au}{{\underline{{A}_{ }}}}                 
\newcommand{\Aul}{{\underline{{A}_{ }}}_{\lambda}}    
\newcommand{\Bul}{{\underline{{B}_{ }}}_{\lambda}}    
\newcommand{\Cul}{{\underline{{C}_{ }}}_{\lambda}}    
\newcommand{\Ma}{{\cal M}}
\begin{document}
\title{Current Trends\\ in Axiomatic Quantum Field Theory\thanks{Talk at 
``Ringberg Symposium on Quantum Field Theory'', 
Ringberg Castle, June 1998}
}
\author{Detlev Buchholz
\\[2mm]
Institut f\"ur Theoretische Physik, Universit\"at 
G\"ottingen,\\ Bunsenstra{\ss}e 9, D-37073 G\"ottingen, Germany\\[0mm]} 
\date{}
\maketitle
\begin{abstract}{\noindent In this article a 
non--technical survey is given of 
the present status of Axiomatic Quantum Field Theory 
and interesting future directions 
of this approach are
outlined. The topics covered are the universal  
structure of the local algebras of observables, 
their relation to the underlying 
fields and the significance 
of their relative positions. 
Moreover, the physical interpretation of the theory 
is discussed with emphasis on 
problems appearing in gauge theories, 
such as the revision of the particle concept, the 
determination of symmetries and statistics from the superselection 
structure, the analysis of the short distance properties and the specific  
features of relativistic thermal states. Some problems appearing in 
quantum field theory on curved spacetimes   
are also briefly mentioned.}
\end{abstract}

\section{Introduction}
\setcounter{equation}{0}
\noindent Axiomatic Quantum Field Theory originated 
from a growing desire in the mid--fifties to have a
consistent mathematical framework for the treatment 
and interpretation of relativistic quantum field theories.
There have been several profound solutions of this   
problem, putting emphasis on different aspects of the theory.
The Ringberg Symposium on Quantum Field Theory 
has been organized in honor of one of the 
founding fathers of this subject, Wolfhart Zimmermann.
It is therefore a pleasure to give an account 
of the present status of the axiomatic approach on this special
occasion.  

It should perhaps be mentioned that the term ``axiomatic''
is no longer popular amongst people working in this 
field since its mathematical connotations have led to misunderstandings. 
Actually, Wolfhart Zimmermann never liked 
it and called this approach
Abstract Quantum Field Theory. 
Because of the modern developments of the subject, 
the presently favored name is  
Algebraic Quantum Field Theory. So the 
invariable abbreviation AQFT seems appropriate in this survey.   

The early successes of AQFT are well known and have been described 
in several excellent monographs \cite{Jo,StWi,BoLoTo}. They led, 
on the one hand, to an understanding of the general mathematical structure
of the correlation functions of relativistic quantum fields and laid the
foundations for the rigorous perturbative and non--perturbative
construction of field theoretic models. On the other hand they  
provided the rules for the physical interpretation of the theory, the most
important result being collision theory and the reduction formulas.  

It was an at first sight perhaps 
unexpected bonus that the precise formulation of the 
foundations of the theory payed off also in other respects. For it led to
the discovery of deep and general features of relativistic quantum
field theories, such as the PCT--theorem, the relation between spin and 
statistics, dispersion relations, bounds on the high energy behavior 
of scattering amplitudes, the Goldstone theorem etc. These results 
showed that the general principles of relativistic quantum field
theory determine a very rigid mathematical framework which 
comprises surprisingly detailed information about the 
physical systems fitting into it. 

These interesting developments changed their direction in 
the seventies for two reasons. First, it became clear  
that, quite generally, the linear and nonlinear properties of the 
correlation functions of quantum fields following from the 
principles of AQFT admit an equivalent Euclidean description 
of the theory in terms of classical random fields. 
Thereby, the construction of relativistic field theories was greatly 
simplified since one could work in a commutative setting \cite{GlJa}. 
Most results in constructive quantum field theory were  
obtained by using this powerful approach which  
also became an important tool in concrete applications. 

The simplification
of the constructive problems outweighed 
the conceptual disadvantage that the Euclidean theory 
does not have a direct physical interpretation.  
To extract this information from the Euclidean formalism  
is frequently a highly non--trivial task and has been the source of 
mistakes. So for the interpretation of the theory 
the framework of AQFT is still indispensable. 

The second impact came from physics. It was the insight 
that gauge theories play a fundamental role in {\em all\/} 
interactions. It was already clear at that time that quantum 
electrodynamics did not fit into 
the conventional setting of AQFT. For the local, covariant 
gauge fields require the introduction of unphysical states 
and indefinite metric. The idea that one could determine from 
such fields in an {\em a priori\/} manner the physical Hilbert 
space had finally to be given up. Features such as the 
phenomenon of confinement in non--Abelian gauge theories 
made it very clear that the specification 
of the physical states is in general a subtle dynamical problem. 

A way out of these problems had already been discovered in the 
sixties, although its perspectives were perhaps   
not fully recognized in the beginning. Namely it became gradually
clear from the structural analysis in AQFT that  
the local observables of a theory 
carry all relevant physical information. In particular, 
the (charged) physical states and their interactions 
can be recovered from them. The situation 
is analogous to group theory, where the set of 
unitary representations can be determined from the abstract structure of the 
group. 

From this more fundamental 
point of view the gauge fields appear to be nothing but 
a device for the construction of the local (gauge invariant) observables 
of the theory in some faithful representation, usually the vacuum 
representation. The determination of the physical states 
and their analysis is then regarded as a problem in representation theory. 

It also became clear that one does not need to
know from the outset the specific physical significance of the local
observables for the interpretation of the theory. All what matters is
the information about their space--time localization properties. From
these data one can determine 
the particle structure, collision cross sections, 
the charges appearing in the theory
and, finally, identify 
individual observables of physical interest, such as the charge and
energy--densities. 

These insights led to the modern formulation of 
AQFT in terms of families (nets)
of algebras of observables which are assigned to the bounded
space--time regions \cite{Ha}. Field operators, even observable ones, 
no longer appear explicitly in this formalism. They are regarded as 
a kind of coordinatization of the local algebras without 
intrinsic meaning; which fields one uses for the 
description of a specific theory is more a matter of convenience
than of principle. 
 
This more abstract point of view 
received, in the course of time, its full justification. First, the framework 
proved to be flexible enough to incorporate non--pointlike localized 
observables, such as the Wilson loops, which became relevant  
in gauge theory. Second, it anticipated to some extent 
the phenomenon of quantum
equivalence, i.e., the fact that certain very differently looking theories,
such as the Thirring model and the Sine--Gordon theory or the recently
explored supersymmetric Yang--Mills theories,    
describe the same physics. The basic insight that  
fields do not have an intrinsic meaning, in contrast to the system of 
local algebras which they generate, found a striking 
confirmation in these examples. Third, the algebraic 
approach proved natural for the discussion of 
quantum field theories in curved spacetime 
and the new types of problems appearing  
there \cite{Wa}. There is also evidence    
that it covers prototypes of 
string theory \cite{Di}. 

So the general framework of AQFT has, for many decades, proved to 
be consistent with the progress in our theoretical understanding
of relativistic quantum physics.
It is the appropriate setting for the discussion of the pertinent 
mathematical structures, the elaboration of methods for 
their physical interpretation, the solution of
conceptual problems and the classification of theories on the 
basis of physical criteria. Some major 
achievements and intriguing open problems in this approach 
are outlined in the remainder of this 
article. 

\section{Fields and algebras}
\setcounter{equation}{0}
As mentioned in the Introduction, 
the principles of relativistic quantum field theory can  
be expressed in terms of field operators and, more generally, 
nets of local algebras. In this section
we give an account of the relation between these two approaches.

We proceed from the for our purposes reasonable  
idealization that spacetime is a classical 
manifold $\Ma$ with pseudo--Riemannian metric $g$.  In the 
main part of this article we assume that $(\Ma,g)$ is 
four--dimensional Minkowski space 
with its standard Lorentzian metric and comment on curved spacetime 
only in the last section. 

\noindent {\em Fields:\/} 
In the original formulation of AQFT 
one proceeds from collections of field operators  
$\phi (x)$ which are assigned to the space--time points $x \in \Ma$, 
\begin{equation} x \longmapsto \{ \, \phi (x) \, \}. \end{equation}
(In order to simplify the discussion we assume that the fields $\phi (x)$ are
observable and omit possible tensor indices.) 
As is well known, this assignment 
requires some mathematical care, it is to be understood in the sense of 
operator--valued distributions. With this precaution in mind the
fundamental principles of relativistic quantum field theory, such as 
Poincar\'e covariance and Einstein causality (locality), can be cast
into mathematically precise conditions on the field operators  
\cite{Jo,StWi,BoLoTo}. As a matter of fact, for given
field content of a theory one can encode these principles into a universal
algebraic structure, the Borchers--Uhlmann algebra of test functions
\cite{Jo}. 

It is evident that such a universal algebra does not contain any
specific dynamical
information. That information can be put in 
by specifying a vacuum state
(expectation functional) on it. This step is the most difficult
task in the construction of a theory. 
Once it has been accomplished, one can extract from 
the corresponding correlation
functions, 
respectively their time--ordered, advanced or retarded counterparts,  
the desired information. 

The fact that the construction of a theory can be accomplished by the 
specification of a vacuum state on some universal algebra has technical
advantages and ultimately led to the Euclidean formulation of
quantum field theory. But it also poses some problems: Given two such
states, when do they describe the same theory? That this is a
non--trivial problem can be seen already in free field theory. There the
vacuum expectation values of the basic free field $\phi_{\, 0}$ 
and those of its Wick power $\phi_{\, 0}^{\, 3}$, say, 
correspond to quite different states on the 
abstract algebra. Nevertheless, they describe the same physics. A less
trivial example, where the identification of two 
at first sight very differently looking 
theories required much more work, can be found in~\cite{Re}.
So in this respect the field theoretic formalism is not intrinsic.\\[3mm]
\noindent {\em Algebras:\/} In the modern algebraic formulation of AQFT
one considers families of $W$*--algebras\footnote{The letter  
$W$ indicates that the respective algebras are closed with respect to weak
limits and * says that they are stable under 
taking adjoints.} $\AO$ of bounded 
operators which are assigned
to the open, bounded space--time regions $\Oo \subset \Ma$,
\begin{equation} \Oo \longmapsto \AO. \label{2.2} \end{equation} 
Each $\AO$ is regarded as the algebra generated by the 
observables which are localized in the region $\Oo$; it is   
called the {\em local algebra\/} affiliated with that region. 
Again, the principles of locality and Poincar\'e
covariance can be expressed in this setting in a straightforward 
manner \cite{Ha}. In addition there holds   
the property of isotony, i.e., 
\begin{equation}
\Al (\Oo_1) \subset \Al (\Oo_2) \quad \mbox{if} \quad 
\Oo_1 \subset \Oo_2. \label{2.3}
\end{equation}
This condition expresses the obvious fact that the set of observables 
increases with the size of the localization region. 
Despite its at first sight almost tautological content, it is this  
net structure (nesting) of the local algebras which comprises the 
relevant physical information about a theory. 
To understand this fact one has to recognize  
that the assignment of algebras to a
given collection of space--time regions will be very different 
in different theories.\\[3mm]
The algebraic version of AQFT defines a conceptually and mathematically
compelling framework of local relativistic quantum physics and has 
proved very useful for the general structural analysis. It is 
rather different, however, from the field theoretic formalism which 
one normally uses 
in the construction of models. The clarification of the relation between 
the two settings is therefore an important issue.\\[3mm]
{\em From fields to algebras:\/} The problems appearing in the
transition from the field theoretic 
setting to the algebraic one are of a similar nature as in the
transition from representations of Lie algebras to representations of 
Lie groups: one has to deal with regularity properties of unbounded 
operators. Heuristically, one would be inclined to 
define the local algebras by appealing
to von Neumann's characterization of concrete $W$*--algebras
as double commutants of sets of operators,   
\begin{equation}
\AO \ = \ \{ \phi (x) \, : \, x \in \Oo \}^{\prime \prime},  
\end{equation} 
that is they ought to be the smallest weakly closed algebras 
of bounded operators 
on the underlying Hilbert space which are generated by the  
(smoothed--out) observable fields in the respective space--time region. 
Because of the subtle properties of unbounded operators it is, 
however, not clear from the outset that the so--defined algebras
comply with the physical constraint of locality assumed in the
algebraic setting. 

The first courageous steps in the analysis of this problem  
were taken by Borchers and Zimmermann 
\cite{BoZi}. They showed that if the vacuum $| 0 \rangle$ is an
analytic vector for the fields, i.e., if the formal power series of
the exponential function of smeared fields, 
applied to $| 0 \rangle$,  
converge absolutely, then the passage from the fields to the local 
algebras can be accomplished by the above formula. Further progress on the 
problem was made in \cite{DrFr}, where it was shown 
that fields satisfying so--called linear energy bounds 
also generate physically acceptable nets of local algebras in 
this way. 

The latter result covers all
interacting relativistic quantum field theories which have been rigorously 
constructed so far. As for the 
general situation, the most comprehensive results are 
contained in \cite{BoYn} and references quoted there. 
In that analysis certain specific  
positivity properties of the vacuum expectation values of fields 
were isolated as crucial pre--requisite for the passage from fields to
algebras. In view of these profound results, it can now safely be 
stated that the algebraic framework is a proper generalization of
the original field theoretic setting.\\[3mm]
{\em From algebras to fields:\/}  As already mentioned, the 
algebraic version of AQFT is more general than the field theoretic one
since it covers also finitely localized observables, such as Wilson loops or
Mandelstam strings, which are not 
built from observable pointlike localized 
fields. Nevertheless, the point--field content of a theory
is of great interest since it includes distinguished
observables, such as currents and the stress energy tensor. 

Heuristically, the point--fields of 
a theory can be recovered from the local algebras by the  
formula 
\begin{equation} 
\{ \,  \phi (x) \,  \} \ = \ \bigcap_{\Oo \, \ni \, x} \, \overline{\AO}\,. 
\end{equation}
It should be noticed here that one would not obtain the desired fields 
if one would simply take the
intersection of the local algebras themselves, 
which is known to consist only of
multiples of the identity. Therefore, one first has to complete the
local algebras in a suitable topology which allows for the appearance
of unbounded operators (respectively forms). This step is 
indicated by the bar. 

The first profound results on this problem were obtained in
\cite{FrHe}, where it has been proposed to complete the local algebras   
with the help of suitable energy norms which are sensitive to the 
energy--momentum transfer of the observables. Using this device,  
it was shown that one can reconstruct from local algebras, 
if they are generated by sufficiently ``tame'' fields, the underlying 
field content by taking intersections as above. 
These results were later refined in various directions \cite{Wo}. 
They show that the step from fields to algebras can be reversed. 

From a general point of view it would, however, be  
desirable to clarify the status of point fields in the algebraic
setting in a more intrinsic manner, i.e., without assuming their existence 
from the outset. An interesting proposal in this direction 
was recently made in \cite{HaOj}. There it was argued that the 
presence of such fields is encoded in phase space properties of 
the net of local algebras\footnote{A quantitative measure of the 
phase space properties of local algebras is
given in Sec.\ 7.} and that the field content can be 
uncovered from the algebras by using notions from sheaf
theory. In \cite{Bos}
this idea was put into a more suitable mathematical form and 
was also confirmed in models. The perspectives of
this new approach appear to be quite interesting. 
It seems, for example, that one can establish in this setting 
the existence of 
Wilson--Zimmermann expansions \cite{WiZi} for products of field operators.  
Such a result would be a major step towards the ambitious goal, 
put forward in \cite{HaOj}, to characterize the dynamics of nets of 
local algebras directly in the algebraic setting. 
\section{Local algebras and their inclusions}
\setcounter{equation}{0}
Because of their fundamental role in the algebraic approach, much work has 
been devoted to the clarification of the 
structure of the local algebras and of their inclusions. We 
cannot enter here into a detailed 
discussion of this subject and only give an  
account of its present status. To anticipate the perhaps most    
interesting perspective of the more recent results: There is evidence that 
the dynamical information of a relativistic 
quantum field theory is encoded in, and can be uncovered from   
the relative positions of a few 
(depending on the number of space--time dimensions) algebras of 
specific type. This insight may be the starting point for a novel 
constructive approach to relativistic quantum field theory. 

To begin, let us recall that the center of a $W$*--algebra 
is the largest sub--algebra of operators commuting with all
operators in the algebra. A $W$*--algebra is called a factor if its center 
consist only of multiples of the identity, and it is said to be 
hyperfinite if it is generated by its finite dimensional sub--algebras. 
The hyperfinite factors have been completely classified, there 
exists an uncountable number of them.

Because of this abundance of different types of algebras it is  
of interest that the local algebras appearing in quantum field
theory have a universal (model--independent) structure \cite{BuDaFr}, 
they are generically isomorphic to the tensor product 
\begin{equation}
\AO \, \simeq \, {\cal M} \otimes {\cal Z}, 
\end{equation}
where ${\cal M}$ is the unique hyperfinite type III$_1$ factor 
according to the classification of Connes, and ${\cal Z}$ is an Abelian
algebra. That the local algebras are hyperfinite is encoded in 
phase space properties of the theory, the type III$_1$ property is
a consequence of the short distance structure, cf.\  \cite{BuDaFr} 
and references quoted there. The possible appearance of a non--trivial
center ${\cal Z}$ in a local algebra 
is frequently regarded as a nuisance, but it cannot be 
excluded from the outset. 

Under the above generic conditions also  
the global (quasilocal) algebra   
\begin{equation}
\Al \, = \, \overline{\bigcup_\Oo \AO},
\end{equation} 
which is the $C$*--inductive limit of all local algebras, 
is known to be universal. It is the so--called 
``proper sequential type I$_\infty$ funnel''.   
So one has very concrete information about the mathematical   
objects appearing in the 
algebraic setting. From the conceptual point of view, 
these results corroborate the 
insight that the individual local algebras as well as the
global one do not comprise any specific physical 
information. This information
is entirely contained in the ``arrow'' in (\ref{2.2}), i.e., the map from 
space--time regions to local algebras. 

In view of these results it is natural to have a closer
look at the possible ``relative positions'' of the local algebras with 
respect to each other. Depending on the location of the regions it 
has been possible to characterize in purely 
algebraic terms the following geometric situations.\\[3mm] 
(a) The closure of $\Oo_1$ is contained in the interior of $\Oo_2$.  
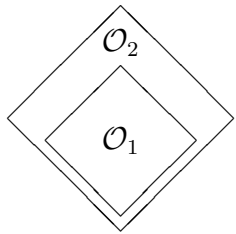
\begin{center}
\setlength{\unitlength}{10mm}
\begin{picture}(6,3)
\put(2,1){\line(1,1){1}}
\put(2,1){\line(1,-1){1}}
\put(4,1){\line(-1,1){1}}
\put(4,1){\line(-1,-1){1}}
\put(2,0.5){\makebox(2,1){$\Oo_1$}}
\put(1.5,1.3){\line(1,1){1.5}}
\put(1.5,1.3){\line(1,-1){1.5}}
\put(4.5,1.3){\line(-1,1){1.5}}
\put(4.5,1.3){\line(-1,-1){1.5}}
\put(2,1.8){\makebox(2,1){$\Oo_2$}}
\end{picture}
\end{center}
\vspace*{2mm}
\noindent In that case there holds generically  
\begin{equation}
\Al ( \Oo_1 ) \, \subset {\cal N} \, \subset \Al ( \Oo_2 ), 
\end{equation}
where ${\cal N}$ is a factor of type I$_\infty$, i.e., an algebra 
which is isomorphic to the algebra of all bounded operators on some 
separable Hilbert space. This ``split property'' of the local 
algebras has been established in all quantum field theories with 
reasonable phase space properties, cf.\ \cite{BuDaFr} and references 
quoted there. It does {\em not} hold if the two regions have common
boundary points \cite{SuWe}. \pagebreak

\noindent (b) $\Oo_1$ and $\Oo_2$ are spacelike separated.  
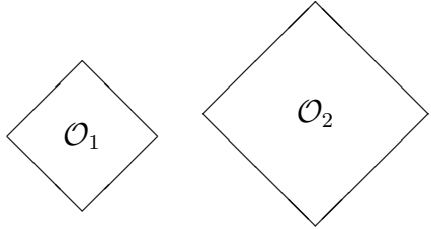
\begin{center}
\setlength{\unitlength}{10mm}
\begin{picture}(6,3)
\put(0,1){\line(1,1){1}}
\put(0,1){\line(1,-1){1}}
\put(2,1){\line(-1,1){1}}
\put(2,1){\line(-1,-1){1}}
\put(0,0.5){\makebox(2,1){$\Oo_1$}}
\put(2.6,1.3){\line(1,1){1.5}}
\put(2.6,1.3){\line(1,-1){1.5}}
\put(5.6,1.3){\line(-1,1){1.5}}
\put(5.6,1.3){\line(-1,-1){1.5}}
\put(3.1,0.8){\makebox(2,1){$\Oo_2$}}
\end{picture}
\end{center}
\vspace*{2mm}
\noindent Then, under the same conditions as in (a), it follows
that the $W$*--algebra which is generated by the two local algebras 
associated with these regions is
isomorphic to their tensor product, 
\begin{equation}
\Al (\Oo_1) \vee \Al (\Oo_2)  \, \simeq \, \Al (\Oo_1) \otimes \Al
(\Oo_2). 
\end{equation}
Thus the local algebras satisfy a condition of causal 
(statistical) independence, 
which may be regarded as a 
strengthened form of the locality postulate.\\[2mm]  
(c) $\Oo_1$ and $\Oo_2$ are wedge--shaped regions, bounded
by two characteristic planes, such that $\Oo_1 \subset \Oo_2$ and 
the edge of $\Oo_1$ is
contained in a boundary plane of $\Oo_2$.
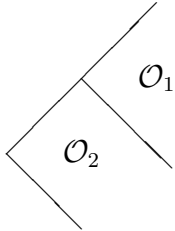
\begin{center}
\setlength{\unitlength}{10mm}
\begin{picture}(4,3)
\put(0,1){\line(1,1){2}}
\put(0,1){\line(1,-1){1}}
\put(0,1){\makebox(2,0){$\Oo_2$}}
\put(1,2){\line(1,-1){1.2}}
\put(1,2){\makebox(2,0){$\Oo_1$}}
\end{picture}
\end{center}
\vspace*{2mm}
\noindent In this geometric situation the corresponding algebras give 
rise to so--called ``half--sided modular inclusions'' or, more
generally, ``modular intersections''.  
It is a striking fact that one can 
reconstruct from a few algebras in this specific position 
a unitary representation of the 
space--time symmetry group, the PCT operator and the net of 
local algebras \cite{Wi1,Wi2}. 
This observation substantiates the claim that the 
dynamical information of a theory is contained in the relative
position of the underlying algebras. 
Thus the concept of modular inclusions and intersections  
seems to be a promising starting point for the direct construction of  
nets of local algebras.

The preceding results rely heavily on modular theory, which  
has become an indispensable tool in the algebraic approach. It is 
not possible to outline here the many interesting applications which are 
based on these techniques. Some recent results and further 
pertinent references can be found in \cite{Bo}. 

\section{Particle aspects}
\setcounter{equation}{0}
We turn now to the physical interpretation of the mathematical 
formalism of AQFT. Here the basic ingredient is the notion of 
particle.  According to Wigner 
the states of a single particle are to be described by
vectors in some irreducible representation of the Poincar\'e group, 
or its covering group. This characterization of particles has been extremely 
useful in the solution of both, constructive and conceptual 
problems. 

As for the interpretation of the theory, the particle concept enters 
primarily in collision theory, whose first precise version was 
given by Lehmann, Symanzik and Zimmermann \cite{LeSyZi}. 
By now collision theory has been rigorously established in 
AQFT, both for massive and
massless particles \cite{Ha}. These results formed the basis for the
derivation of analyticity, crossing and growth properties of the scattering 
amplitudes \cite{MaCh}. There remain, however, many open problems in 
this context. In view of the physical relevance of 
gauge theories it would, for example, be desirable to 
determine the general properties of scattering amplitudes of 
particles carrying a gauge charge. In physical gauges such particles 
require a description in terms of non--local fields, hence the classical
structural results cannot be applied. Some remarkable progress on this 
difficult problem has been presented in \cite{BrEp}. 

Another longstanding question   
is the problem of asymptotic completeness (unitarity of 
the S--matrix). Even in the models which have been rigorously 
constructed, a complete solution of this problem is not known \cite{Ia}. 
So the situation is quite different from quantum mechanics, where 
the problem of asymptotic completeness 
was solved in a general setting almost two decades ago. 

In order to understand the origin of these difficulties,   
one has to realize that, in contrast to quantum mechanics, 
one deals in quantum field theory with systems
with an infinite number of degrees of freedom. In such systems there
can occur the formation of superselection sectors which require 
an extension of the original Hilbert space.
So one first has to determine in a theory 
the set of all superselection sectors and particle types  
before a discussion of the problem of 
asymptotic completeness becomes meaningful. 
In models this step can sometimes be 
avoided by technical assumptions, such as restrictions on the 
size of coupling constants, by which the formation of    
superselection sectors and new particles can be excluded. But the 
determination of the full physical Hilbert space from the 
underlying local operators is an inevitable step 
in any general discussion of the problem. Some
progress on this problem will be outlined below. 

Still another important problem which deserves mentioning here  
is the treatment of particles carrying charges of
electromagnetic type. As is well known, the states of  
such particles cannot consistently be described in the way proposed by 
Wigner, cf.\ \cite{Bu1} and references quoted there. 
In the discussion of scattering processes involving   
such particles this problem can frequently be circumvented 
by noting that an infinite number of soft massless particles 
remains unobserved. Because of this fact one can  
proceed to an ``inclusive description'', where 
the difficulties disappear. This trick obscures, 
however, the specific properties of the electrically charged particles.   
It seems therefore worthwhile to analyze their uncommon  
features which may well be accessible to experimental tests. 

It is apparent that progress on these problems 
requires a revision of the particle concept. A proposal in this
respect which is based on Dirac's idea of improper states of sharp
momentum has been presented in \cite{Bu2}. The common mathematical treatment 
of improper states as vector--valued distributions would not work in
the general setting since it assumes the superposition principle which is
known to fail for momentum eigenstates carrying electric charge.  
Instead one defines the improper states as linear 
maps from some space (more precisely, left ideal) 
of localizing operators $L \in \Al$ into the physical Hilbert space, 
\begin{equation}
L \longmapsto L \, | \sp, \iota \rangle.
\end{equation}
Here $\sp$ is the momentum of the particle and $\iota$ subsumes its
charges, mass and spin. In quantum mechanics, the simplest example of 
a localizing operator $L$ which transforms the ``plane waves''
$ | \sp, \iota \rangle$ into normalizable
states is the projection onto a compact region of configuration space.
In quantum field theory such localizing operators 
can be constructed out of local operators by
convolution with suitable test functions which restrict the 
energy--momentum transfer of the operators to spacelike values. 
It can be shown that by acting with any such operator 
on an improper momentum eigenstate of
a particle one obtains a Hilbert space vector.   
Thus, from a mathematical point of view, the improper states are
weights on the algebra of observables $\Al$. 

Using this device one can, on the one hand, determine the  
particle content of a theory from the states $| \Phi \rangle$ in the 
vacuum sector by means of the formula \cite{Bu2}
\begin{equation}
\lim_{t \rightarrow \pm \infty} \int \dx \,
\langle \Phi | \alpha_{t,\ssx} (L^* L) | \Phi \rangle \, = \,
\sum_\iota \int d\mu_\pm (\sp,\iota) \,
\langle \sp, \iota | L^* L | \sp, \iota \rangle. \label{4.2}
\end{equation}
Here $L$ is any localizing operator, ${\alpha}_{t,{\ssx}}$  are  
the automorphisms inducing the  
space--time translations on $\Al$ and $\mu_\pm$ are measures depending
on $| \Phi \rangle$. Thus if one analyzes the states $| \Phi \rangle$  
at asymptotic times by spatial averages of localizing operators, they
look like mixtures of improper single particle states. These mixtures are   
formed by the members of the incoming respectively outgoing particle 
configurations in the state $| \Phi \rangle$ which 
generically include also pairs of oppositely 
charged particles. By decomposing the 
mixtures in (\ref{4.2}) one can therefore 
recover all particle types.
As outlined in \cite{BuPoSt}, this result also allows one to 
recover from the underlying local observables in the vacuum 
sector the pertinent physical
Hilbert space of the theory.

Relation (\ref{4.2}) does not only establish a method for the
determination of the particles in a theory, but it also provides 
a framework for their general analysis. It is of interest that this 
framework also covers  
particles carrying charges of electromagnetic type. Whereas for a
particle of Wigner type the corresponding improper states lead, 
after localization, to vectors in the same sector of the physical 
Hilbert space, this
is no longer true for electrically charged particles. There one finds
that the vectors $L | \sp, \iota \rangle$ and 
$L^\prime | \sp^\prime, \iota \rangle$ are orthogonal for any choice
of localizing operators $L, L^\prime$ if the momenta 
$\sp,\sp^\prime$ are different. So the superposition 
principle fails in this case and wave  packets of improper states 
cannot be formed. 
Nevertheless, the charges, mass and spin
of such particles can be defined and have the values found by 
Wigner in his analysis. The only possible exception are massless
particles whose helicity need not necessarily be restricted to
(half--)integer values \cite{Bu2}. 

So there is progress in our general understanding of the particle
aspects of quantum field theory. Further advancements  
seem to require, however, new methods such as 
a more detailed harmonic analysis of the space--time automorphisms \cite{Bo1}.
\section{Sectors, symmetries and statistics}
\setcounter{equation}{0}
One of the great achievements of AQFT is the 
general understanding of the structure of superselection sectors in 
relativistic quantum field theory, its relation to the appearance 
of global gauge groups and the origin and classification of statistics. 
Since this topic has been expounded in the monographs 
\cite{Ha,BaWo} we need to mention here only 
briefly the main results and open problems.

The superselection sectors of interest in quantum
field theory correspond to specific 
irreducible representations of the algebra of observables $\Al$, 
more precisely, to their respective unitary equivalence classes. 
It is an important fact that each sector has representatives
which are (endo)morphisms $\rho$ of 
the algebra $\Al$ (or certain canonical extensions of it), so there holds 
\begin{equation}
\rho (\Al) \subset \Al. 
\end{equation}
Using this fact  
one can distinguish various types 
of superselection sectors according to the localization properties 
of the associated morphisms.\\[2mm]
{\em Localizable Sectors:\/} These sectors 
have been extensively studied by Doplicher, Haag and Roberts. Each such 
sector can be characterized by the property that for 
any open, bounded space--time region $\Oo \subset \Ma$ 
there is a morphism $\rho_\Oo$, 
belonging to this sector, which is localized in $\Oo$ in the sense that
it acts trivially on the observables 
in the causal complement $\Oo^{\, \prime}$ of $\Oo$, 
\begin{equation}
\rho_\Oo \rest \Al (\Oo^{\, \prime}) = \mbox{id}. 
\end{equation}
Localizable sectors describe charges, such as baryon number, 
which do not give rise to long range effects.\\[2mm] 
{\em Sectors in Massive Theories:\/} It has been shown in 
\cite{BuFr} that the superselection 
sectors in quantum field theories describing massive particles can
always be represented by morphisms $\rho_{\, \cal C}$ which are localized 
in a given spacelike cone ${\cal C}$, 
\begin{equation}
\rho_{\, \cal C} \rest \Al ({\cal C}^{\, \prime}) = \mbox{id}. 
\end{equation} 
This class of sectors includes also  
non--localizable charges appearing in certain massive gauge theories. 
The preceding result says that the long range effects of such charges can 
always be accommodated in extended string--like regions.\\[2mm]  
\indent It is possible to establish for both classes of sectors under very
general conditions, the most important one being a maximality
property of the underlying algebras of local observables (Haag--duality),
the existence of a composition law. Namely,   
if $\iota, \kappa, \lambda$ are labels characterizing the 
sectors there holds for the corresponding morphisms the relation
\begin{equation}
\rho_\iota \, {\scriptstyle \circ} \, 
\rho_\kappa = \sum_\lambda c(\iota, \kappa, \lambda)
\ \rho_\lambda, 
\end{equation} 
where $c(\iota, \kappa, \lambda)$ are integers and the summation is to be
understood in the sense of direct sums of representations. Moreover,
for each $\rho_\iota$ there is a charge conjugate morphism
$\overline{\rho}_\iota$ such that $\rho_\iota \, {\scriptstyle \circ}
\, \overline{\rho}_\iota$ contains the identity. 

It is a deep result
of Doplicher and Roberts that these general facts imply the existence of 
charged fields $\psi$ of Bose or Fermi type which 
connect the states in the
various superselection sectors. More precisely,  
each morphism $\rho$ can be represented in the form 
\begin{equation}
\rho \, (\, \cdot \,) = \sum_{m=1}^d \psi_m \, \cdot \, \psi_{m}^* \, ,
\end{equation} 
where the fields $\psi_m$ have the same localization properties 
as $\rho$ and satisfy Bose or Fermi commutation relations at 
spacelike distances. Moreover, these fields transform as 
tensors under the action of some compact group $G$, 
\begin{equation}
\gamma_g \, (\psi_n) = \sum_{m=1}^d D_{n \, \!  m}(g^{-1}) \, \psi_m \, ,
\end{equation}  
where $\gamma_{\, \cdot}$ are the automorphisms inducing this action and 
$D_{n \, \! m} ( \cdot )$ is some $d$--dimension\-al representation of $G$. 
The observables are exactly the fixed points under 
the action of $G$ and the whole structure is uniquely determined by 
the underlying sectors. 

In order to appreciate the strength of this result one has to notice  
that it does not hold in low space--time dimensions. In that case
there can occur 
fields with braid group statistics and the superselection structure
can in general not be described by the representation theory of 
compact groups, while more  
complex symmetry structures, such as ``quantum groups'', seem to
emerge \cite{FrReSch}. These facts have 
stimulated much work in the general analysis of low dimensional
quantum field theories in recent years, yet we cannot comment here
on these interesting developments. 

With regard to physical space--time, there are several interesting
problems which should be mentioned here. First, it would
be of interest to understand in which way the presence of
supersymmetries is encoded in the superselection structure
of a theory. It seems that this problem has not yet been 
thoroughly studied in the general framework of AQFT.  

Second, there is the longstanding  
problem of the  superselection structure in theories with long range 
forces, such as Abelian gauge theories with unscreened charges of 
electromagnetic type. In such theories there exists for each value 
of the charge an abundance of sectors due to the 
multifarious ways in which accompanying clouds of low energy massless 
particles can be formed. These clouds obstruct the general analysis of the  
superselection structure since it is difficult to disentangle their 
fuzzy localization properties from those of the charges one is actually 
interested in. 

A promising step towards the solution of this problem
is the observation, made in some models \cite{BuDoMoRoSt}, that 
charges of electromagnetic type have certain clearcut 
localization properties in spite of their 
long range effects. 
Roughly speaking, they appear to be localized in a given 
Lorentz system with respect to distinguished observables, such as  
current densities. In the framework of AQFT this restricted
localizability of sectors can be expressed by assuming that 
there is a subnet $\Oo \longmapsto \BO \subset \AO$ on which 
the corresponding morphisms are localized, 
\begin{equation}
\rho_\Oo \rest \BOp = \mbox{id}. 
\end{equation}
In contrast to the preceding 
class of localizable sectors, the net $\Oo \longmapsto \BO$  
need not be maximal, however. In the case of electrically  
charged sectors, it will also not be stable under Lorentz transformations. 
So the general results of Doplicher and Roberts cannot be applied 
in this case. Yet there is evidence \cite{BuDoMoRoSt}
that if the localizing subnet of observables is sufficiently big 
(it has to satisfy a condition of 
weak additivity) one can still establish symmetry and statistics 
properties for the physically interesting class of so--called simple sectors 
which are induced by automorphisms $\rho$ of the algebra of observables
with the above localization properties. 

Another important issue which is closely related to the sector 
structure is the problem of symmetry breakdown. The consequences of the 
spontaneous breakdown of internal symmetries are well understood 
in AQFT in the context of localizable sectors, cf.\ 
\cite{BuDoLoRo} and references quoted there. They lead to 
a degeneracy of the vacuum state and, 
under more restrictive assumptions, to the appearance 
of massless Goldstone particles. 
For the class of cone--like
localizable sectors the consequences of spontaneous
symmetry breaking are less clear, however. One knows from model studies 
in gauge theories that under these circumstances 
there can appear a mass gap in the theory (Higgs mechanism).  
But the understanding of this phenomenon in the general framework 
of AQFT is not yet in a satisfactory state. 
\section{Short distance structure}
\setcounter{equation}{0}
Renormalization group methods have proved to be a powerful tool 
for the analysis of the short distance (ultraviolet) properties 
of field--theoretic models. They provided the basis for 
the discussion of pertinent physical concepts, such as the notion of parton, 
confinement, asymptotic freedom, etc. In view of these successes 
it was a natural step to transfer these methods to the abstract  
field theoretic setting \cite{Zi}. More recently, the method has   
also been established in the algebraic formulation of 
AQFT \cite{BuVe}. We give here a brief account of the latter 
approach in which renormalization group transformations are introduced in 
a novel, implicit manner. 

The essential idea is to consider functions $\Au$
of a parameter $\lambda \in \RR_+$, fixing the space--time scale, 
which have values in the algebra of observables, 
\begin{equation}
\lambda \longmapsto \Aul \in \Al.
\end{equation}
These functions form under the obvious pointwise defined algebraic operations 
a normed algebra, the scaling algebra $\Alu$, on which the 
Poincar\'e transformations $(\Lambda,x)$ act continuously by 
automorphisms $\underline{\alpha}_{\, \Lambda, x}$ according to 
\begin{equation}
\left( {\underline{\alpha}}_{\, \Lambda, x} \, {(\Au)} \right)_\lambda
\doteq \alpha_{\, \Lambda, \lambda x} \, (\Aul).  
\end{equation}
The local structure of the original net can be lifted to
$\Alu$ by setting 
\begin{equation}
\AOu \doteq \{ \Au : \Aul \in \Al  (\lambda \Oo), \ \lambda \in
\RR_+ \}.  
\end{equation}
It is easily checked that with these definitions one obtains a 
local, Poincar\'e covariant net of subalgebras of the scaling algebra 
which is canonically associated 
with the original theory. 

We refer the reader to \cite{BuVe} for a discussion of 
the physical interpretation of this formalism and only mention here that 
the values $\Aul$ of the functions $\Au$ are to be regarded as 
observables in the theory at space--time scale $\lambda$. So the 
graphs of the functions $\Au$ establish a relation between observables
at different space--time scales, in analogy to renormalization group 
transformations. Yet, in contrast to the field theoretic setting,  
one need not identify specific observables in the algebraic 
framework since the physical information is contained in the net 
structure. For that reason one has much more freedom in the choice 
of the functions $\Au$  which have to satisfy only the few general 
constraints indicated above. Thus one arrives at a universal framework 
for the discussion of the short distance structure in quantum field 
theory. 

The physical states, such as the vacuum $| 0 \rangle$, 
can be analyzed at any scale with the help of the scaling
algebra and one can define a short distance (scaling) limit 
of the theory by the formula
\begin{equation}
\ \langle 0 | \, \underline{A_{}}_0 \, \underline{B_{}}_0 \cdots 
\underline{C_{}}_0 \, | 0 \rangle \ \doteq \
\lim_{\lambda \rightarrow 0} \ \langle 0 | \, 
\Aul \, \Bul \cdots \Cul \, | 0 \rangle. \label{6.4}
\end{equation}
To be precise, the convergence on the right hand side may only hold for
suitable subsequences of the scaling parameter, and it should also be
noticed that the limit may not be interchanged with the expectation value.
The resulting correlation functions determine a pure vacuum state on the
scaling algebra. So by the reconstruction theorem one obtains a net  
$\Oo \mapsto \Al_0 (\Oo)$ of local algebras and automorphisms 
$\alpha^{(0)}_{\Lambda,x}$ which induce the Poincar\'e transformations
on this net. This scaling limit net describes the properties of 
the underlying theory at very small space--time scales. 

Based on these results the possible structure of scaling limit
theories has been classified in the general framework of 
AQFT \cite{BuVe}. There appear three qualitatively different cases 
(classical, quantum and degenerate limits). There is evidence that 
they correspond to the various possibilities in the field 
theoretic setting of having (no, stable, unstable) 
ultraviolet fixed points. One can also characterize in 
purely algebraic terms those local nets which ought 
to correspond to asymptotically 
free theories. Yet a fully satisfactory clarification of the relation between 
the field theoretic renormalization group and its algebraic version 
requires further work.  

The framework of the scaling algebra has also shed new light on the
physical interpretation of the short distance properties of 
quantum field theories \cite{Bu3}. Particle like entities and 
symmetries appearing only at very small space--time scales, such 
as partons and color, can be
uncovered from a local net of observables by proceeding to its 
scaling limit. Since the resulting net has all properties 
required in AQFT, the methods and results outlined in the preceding 
two sections can be applied to determine these structures.
It is conceptionally very satisfactory that one does
not need to rely on unphysical quantities in this analysis, 
such as gauge fields; moreover, the method has also proven to be
useful as a computational tool.

There remain, however, many interesting open problems in this
setting. For example, one may hope to extract in the scaling 
limit information about the presence of a {\em local\/}
gauge group in the underlying theory. A possible strategy could 
be to consider the scaling limit of the theory with respect 
to base points $p \neq 0$ in Minkowski space and to introduce some  
canonical identification of the respective nets. One may then study how this 
identification lifts to the corresponding algebras of charged fields, 
which are fixed by the reconstruction theorem of Doplicher and Roberts
outlined in the preceding section. It 
is an interesting question whether 
some non--trivial information about the presence
of local gauge symmetries is encoded in this structure. 

Another problem of physical interest is the development of 
methods for the determination of the particle content of a given 
state at short distances which, in contrast 
to the determination of the particle content of a given theory, has not 
yet been accomplished. In order to understand this problem one has
to notice that there holds for any physical state $| \Phi \rangle$
\begin{equation}
\lim_{\lambda \rightarrow 0} \ \langle \Phi | \, 
\Aul \, \Bul \cdots \Cul \, | \Phi \rangle \, = \, 
\, \langle 0 | \, \underline{A_{}}_0 \, \underline{B_{}}_0 \cdots 
\underline{C_{}}_0 \, | 0 \rangle, 
\end{equation}
so all states look like the vacuum state in the scaling limit \cite{BuVe}. 
In order to extract more detailed information
about the short distance structure of $| \Phi \rangle$, 
one has to determine also 
the next to leading contribution of the matrix element
on the left hand side of this relation. One may expect that this term 
describes the particle like structures which appear in the 
state $| \Phi \rangle$ at small scales, but this question has not yet
been explored.  
\section{Thermal states}
\setcounter{equation}{0}
The rigorous analysis of thermal states in 
non--relativistic quantum field theory is an old subject  
which is well covered in the literature. Interest in  
thermal states in relativistic quantum field theory arose only more   
recently. For this reason there is a backlog in our understanding
of the structure of thermal states in AQFT and one may therefore 
hope that this physically relevant 
topic will receive increasing attention in the future. 

We recall that, from the algebraic point of view, a theory is fixed by
specifying a net of local algebras of observables. 
The thermal states correspond 
to distinguished positive, linear and normalized functionals 
$\langle \, \cdot \, \rangle$ on the corresponding global 
algebra $\Al$. We will primarily discuss here 
the case of thermal {\em equilibrium\/} states, which can be characterized in
several physically meaningful ways, the technically most convenient
one being the KMS--condition, which imposes analyticity 
and boundary conditions on the correlation functions \cite{Ha}. 
Having characterized the equilibrium states, it is natural
to ask which properties of a local net matter if such states are to
exist. The important point is that, in answering this question, one
arrives at criteria which distinguish a physically reasonable 
class of theories and can be used for their further analysis. 

It has become clear by now that phase space properties, which
were already mentioned at various points in this article, 
are of vital importance in this context. 
A quantitative measure of these properties, where the
relation to thermodynamical considerations is particularly
transparent, has been
introduced in \cite{BuWi}. In that approach one considers for any   
$\beta > 0$ and bounded space--time region $\Oo$ the linear map 
$\Theta_{\beta, \Oo}$ from the local algebra $\AO$ to the vacuum 
Hilbert space ${\cal H}$ given by 
\begin{equation}
\Theta_{\beta, \Oo} (A) \, \doteq \, e^{- \beta H} A | 0 \rangle,
\label{7.1}
\end{equation}
where $H$ is the Hamiltonian. Roughly speaking, this map
amounts to restricting the operator $e^{- \beta H}$
to states which are localized in $\Oo$, whence the sum  
of its eigenvalues yields the 
partition function of the theory for spatial volume 
$| \Oo |$ and inverse temperature $\beta$.
This idea can be made precise by noting that 
for maps between Banach spaces, such as 
$\Theta_{\beta, \Oo}$, one can introduce a nuclear norm 
$|| \, \cdot \, ||_1$ which is the appropriate 
generalization of the concept of the trace of Hilbert 
space operators. Thus $|| \Theta_{\beta, \Oo} ||_1$ takes the 
place of the partition function 
and provides the desired information on 
the phase space properties (level density) of the theory. 
If the theory is to have reasonable
thermodynamical properties there must hold for small $\beta$ and
large $| \Oo |$ 
\begin{equation}
|| \Theta_{\beta, \Oo} ||_1 \, \leq \, e^{\, c  \, |\Oo|^m 
\, \beta^{-n}}, \label{7.2}
\end{equation}
where $c,m$ and $n$ are positive constants \cite{BuWi}. This
condition, which can be checked in the vacuum sector ${\cal H}$,  
should be regarded as a selection criterion for theories of 
physical interest.

It has been shown in \cite{BuJu} that any local net which satisfies condition 
(\ref{7.2}) admits thermal equilibrium states 
$\langle \, \cdot \, \rangle_\beta$ for all $\beta > 0$. As a
matter of fact, these states can be reached from the vacuum sector
by a quite 
general procedure. Namely, there holds for $A \in \Al$
\begin{equation}
\langle \, A \, \rangle_\beta \, = \, \lim_{\Oo \rightarrow \RR^4} \,
 \mbox{\large $\frac{1}{Z_\Oo}$} 
\ \mbox{Tr} \, E_{\Oo} \, e^{- \beta H} E_{\Oo} \, A,
\end{equation}
where $E_{\Oo}$ projects onto certain ``local'' subspaces of 
${\cal H}$ and $Z_\Oo$ is a normalization constant \cite{BuJu}. This formula   
provides a direct description of the Gibbs ensemble in the 
thermodynamic limit which does not require the 
definition of the theory in a ``finite box''. It has led 
to a sharpened characterization of thermal 
equilibrium states in terms of a {\em relativistic\/} KMS--condition 
\cite{BrBu}. This condition states that for any $A,B \in \Al$ the correlation 
functions  
\begin{equation}
x \longmapsto 
\langle \, A \, \alpha_x (B)\, \rangle_\beta 
\end{equation}
admit an analytic continuation in all space--time variables $x$ into 
the complex tube $\RR^4 + i \, {\cal C}_{\beta}$, where 
${\cal C}_{\beta} \subset V_+$ 
is a double cone of size proportional 
to $\beta$, 
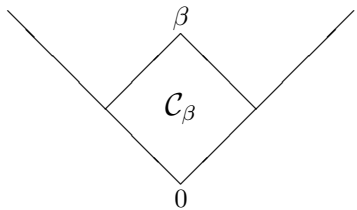
\begin{center}
\setlength{\unitlength}{10mm}
\begin{picture}(6,3.1)
\put(2,1.5){\line(1,1){1}}
\put(2,1.5){\line(1,-1){1}}
\put(4,1.5){\line(-1,1){1}}
\put(4,1.5){\line(-1,-1){1}}
\put(2,1){\makebox(2,1){${\cal C}_{\beta}$}}
\put(2,1.5){\line(-1,1){1.3}}
\put(4,1.5){\line(1,1){1.3}}
\put(2,2.2){\makebox(2,1){\footnotesize ${\beta}$}}
\put(2,-0.2){\makebox(2,1){\footnotesize ${0}$}}
\end{picture}
\end{center}
and the boundary values of these functions at the upper tip of
${\cal C}_{\beta}$ coincide with 
\begin{equation}
x \longmapsto 
\langle \, B \, \alpha_x (A)\, \rangle_\beta.  
\end{equation}
The relativistic KMS--condition may be regarded as a 
generalization of the relativistic spectrum condition to the case of 
thermal equilibrium states. One recovers from it the 
spectrum condition for the vacuum sector in the limit 
$\beta \rightarrow \infty$.

The (relativistic) KMS--condition and the condition of locality lead 
to enlargements of the domains of analyticity of correlation functions 
of pointlike localized fields, in
analogy to the case of the vacuum. Moreover, there exists an analogue 
of the K\"all\'en--Lehmann representation 
for thermal correlation functions which provides a basis for the
discussion of the particle aspects of thermal states. These  
results are, however, far
from being complete. We refer to \cite{BrBu1} for a more detailed
account of this topic and further references. 

Amongst the many intriguing problems in thermal AQFT 
let us also mention the unclear status 
of perturbation theory \cite{St}, the still pending      
clarification of the relation between the Euclidean and Minkowski 
space formulation of the theory \cite{Fr} and the characterization 
of non--equilibrium states. Any progress on these issues would be an  
important step towards the consolidation of thermal quantum field
theory. 
\section{Curved spacetime}
\setcounter{equation}{0}
The unification of general relativity and quantum theory 
to a consistent theory of quantum gravity 
is an important issue which has become a major field of activity in 
theoretical physics. There are many stimulating proposals which 
are based on far--reaching theoretical ideas and novel mathematical 
structures. Yet the subject is still in an experimental state and has
not yet reached a point where one could extract from the various
approaches and results a consistent mathematical formalism with a
clearcut physical interpretation. In a sense, one may compare the
situation with the status of relativistic quantum field theory
before the invention of AQFT.

In view of these theoretical uncertainties and lacking experimental
clues it seems appropriate to treat, in an intermediate step, the
effects of gravity in quantum field theory as a classical
background. This idea has motivated the formulation of AQFT on curved
space--time manifolds $(\Ma,g)$. 
As far as the algebraic aspects are concerned, this step does not
require any new ideas. One still deals with nets (\ref{2.2}) of local
algebras which are assigned to the bounded space--time regions of 
$\Ma$. On these nets there act the  isometries of $(\Ma,g)$ by
automorphisms and
they satisfy the principle of locality (commutativity of observables
in  causally disjoint regions). A novel difficulty which appears in
this setting is the characterization of states  of physical
interest. For in general the isometry group of $(\Ma,g)$ does not
contain global future directed Killing vector fields which could be
interpreted as time translations and would allow for the
characterization of distinguished ground states, representing the
vacuum, or thermal equilibrium states. 

The common strategy to overcome these difficulties is to
invent local regularity conditions which distinguish
subsets (folia) of physically acceptable 
states amongst the set of all states on 
the global algebra of observables. Such 
conditions have successfully been formulated for free field theories
\cite{Wa}, yet their generalization turned out to be
difficult and required new ideas. 
There are two promising proposals which can be applied to 
arbitrary theories, the ``condition of local stability'' 
\cite{HaNaSt}, which fits well into the algebraic setting, and 
the ``microlocal spectrum condition'' \cite{Ra,BrFrKo}, which 
so far has only been stated in a setting based on point fields. 
These conditions have proved to be useful for the discussion of
prominent gravitational effects, such as the Hawking temperature 
\cite{HaNaSt,FrHa}, and
they provided the basis for the formulation of a consistent
renormalized perturbation theory on curved spacetimes \cite{BrFr}.

In the long run, however, it appears to be inevitable to solve the
problem of characterizing states describing specific physical
situations. If the underlying spacetime is sufficiently symmetric in
the sense that it admits Killing vector fields which are future
directed on subregions of the space--time manifold (an example being de
Sitter space) one can indeed identify vacuum--like states by
symmetry and local stability properties \cite{BrEpMo,BoBu}. The
general situation is unclear, however. It has been suggested in 
\cite{BuDrFlSu} 
to distinguish the physically preferred states by
a ``condition of geometric modular action'' which can be
stated for a large class of space--time manifolds. But this  proposal
has so far been proven to  work only for spacetimes where the
preceding local stability 
conditions are also applicable. So there is still much work needed 
until our understanding of this
important issue may be regarded as satisfactory. 

Even more mysterious is the generalization of the particle concept to
curved space--time manifolds and the description of collision
processes. Since curvature gives rise to interaction with the
classical background its effects have to be taken into 
account in the characterization  
of the corresponding states. These problems are of a similar
nature as those occuring 
in the description of particle states in Minkowski space in
the presence of long range forces or in thermal states. One may 
therefore hope that progress in the understanding of the latter
problems will also provide clues to the solution of these  
conceptual difficulties in curved spacetime. 
\section{Concluding remarks}
\setcounter{equation}{0}
In the present survey of AQFT 
emphasis was put on issues which are of relevance for the
discussion of relativistic quantum field theory in physical
spacetime. There has been considerable progress in recent years in our
understanding of the general mathematical framework and its physical
interpretation. In particular, it has become clear that the modern  
algebraic approach is suitable for the discussion of the   
conceptual problems appearing in 
gauge theories. Yet there are still many intriguing questions which
deserve further clarification. 

Several topics which are presently in the
limelight of major research activities had to be omitted here, such as
the general analysis of low--dimensional quantum field theories.
These theories provide a laboratory for the exploration 
of new theoretical ideas and methods and their 
thorough investigation brought to light 
novel mathematical structures which stimulated the interest of 
mathematicians. It was also not possible to outline here 
the many pertinent results and interesting   
perspectives which are based on the powerful techniques of 
modular theory. For an account of
these exciting developments we refer to \cite{Ka,Sch}.

Thus, in spite of its age, AQFT is still very much alive and 
continues to be a valuable source of our theoretical understanding. 
One may therefore hope that it will eventually lead, 
together with the constructive efforts, 
to the rigorous consolidation of relativistic quantum field theory. 

\end{document}